\documentclass{article}

\usepackage{arxiv}

\usepackage[utf8]{inputenc}
\usepackage[T1]{fontenc}
\usepackage{amsmath,amssymb,amsfonts}
\usepackage{graphicx}
\usepackage{booktabs}
\usepackage{array}
\usepackage{url}
\usepackage{microtype}
\usepackage[hidelinks]{hyperref}
\graphicspath{{./figs/}}

\newcommand{\ci}[2]{[#1, #2]}

\title{Same Name, Different Server: A Security Census of Silent Drift in the Model Context Protocol Ecosystem}

\author{
 Obada Kraishan \\
  College of Media and Communication\\
  Texas Tech University\\
  Lubbock, TX 79409, USA \\
}

\begin{document}
\maketitle

\begin{abstract}
The Model Context Protocol (MCP) has become the common interface through which large language model applications reach external tools, and its public registry now distributes thousands of community-built servers with little of the vetting infrastructure that mature package ecosystems have accumulated. This paper reports a census of that ecosystem. We harvested the full public MCP registry (21{,}643 servers, 72{,}606 version records, August 2026 snapshot), fetched source code for 14{,}353 servers, and applied a pattern-based scanner covering an eight-class threat catalogue whose accuracy we measured against 414 hand-labeled findings. Observed prevalence is dominated by unauthenticated network exposure (9.57\% of scanned servers); after correcting each class by its measured precision, 11.14\% observed high-severity prevalence reduces to roughly 7.6\%. The central finding concerns instability rather than any single weakness: 51.1\% of multi-version servers changed what they advertise between versions, 40.6\% did so silently, and 4.2\% redirected their remote endpoint to a different host while keeping their registry identity, a change the protocol never surfaces to installed clients. Silent drift is associated with nearly threefold higher odds of a high-severity finding (OR = 2.96, 95\% CI [2.56, 3.42]). Popularity offers only weak protection (OR = 0.78 per unit of log stars), so star counts are a poor proxy for safety. We derive concrete recommendations for registry design, client-side pinning, and scanner triage, and release an anonymized artifact.
\end{abstract}


\section{Introduction}
When a language model agent edits a file, queries a database, or books a meeting, it increasingly does so through a Model Context Protocol (MCP) server. Introduced by Anthropic in late 2024 \cite{anthropic2024mcp}, MCP standardizes how model applications discover and invoke external tools, and it has been adopted across the major model providers and developer platforms. A public registry now lets anyone publish a server under a name that clients resolve at install time. That convenience creates a supply chain: the code behind a registry name runs with the user's credentials, network position, and file system, yet the ecosystem is young enough that the guardrails other package registries developed over a decade, such as signing, provenance attestation, and automated malware review, are largely absent.

Security discussion around MCP has so far been driven by demonstrations. Tool poisoning, in which hidden instructions in tool metadata steer the model \cite{invariant2025poisoning}, confused-deputy token forwarding, and rug pulls, in which a benign server later turns hostile, have all been shown feasible in the lab \cite{radosevich2025audit, zhao2025attack, hou2025landscape}. What such demonstrations cannot answer is a quantitative question: how common are these weaknesses in the population of servers users actually install, and which properties of a server predict them? Without population-level numbers, defenders cannot prioritize, registry operators cannot tell which structural gaps matter, and researchers cannot tell whether the next attack paper describes a widespread condition or a corner case.

This paper answers that question with a census rather than a sample. We harvested every entry in the public MCP registry at a single snapshot (August 2026): 21{,}643 servers and 72{,}606 version records. For the 79.4\% of servers declaring a source repository we fetched code from GitHub, yielding 13{,}051 repositories, 248{,}817 files, and a scan population of 14{,}353 servers. A pattern-based scanner covering eight threat classes produced the security measurements; a version-differencing pipeline over 8{,}900 multi-version servers produced the stability measurements; and 414 hand-labeled findings anchor every prevalence figure to a measured precision, so the paper reports what the scanner sees and how much of it is real.

Three research questions structure the study, each deliberately single-barreled:

\begin{itemize}
\item \textbf{RQ1.} How prevalent is each threat class across the public MCP server population?
\item \textbf{RQ2.} How often do servers silently change what they advertise across published versions?
\item \textbf{RQ3.} Which observable server characteristics predict security posture?
\end{itemize}

RQ1 includes, as a subsection, the delegation-risk classes (ambient authority and token passthrough), which merit joint treatment because both break the assumption that a tool acts only within a scope the user granted. RQ2 additionally tests whether instability relates to security posture, and that relation turns out to be the study's most consequential result. Scanner accuracy is treated as validation inside the Method, not as a research question: its numbers qualify everything else.

The main findings are as follows. First, weaknesses are common but concentrated: 15.2\% of scanned servers carry at least one finding and 11.1\% at least one high-severity finding, with unauthenticated network exposure alone accounting for most of the high-severity mass; after precision correction the high-severity picture shrinks but does not disappear. Second, the registry is unstable in a way its own data model cannot express: half of multi-version servers changed their advertised description, packages, or endpoints between versions, two in five did so without any identifier changing, and 370 servers moved their remote endpoint to a different host while their registry name stayed fixed, the closest observable analogue of a rug pull. Servers with silent drift have 2.96 times the odds of a high-severity finding. Third, popularity is a weak signal: rank correlations between stars and findings are indistinguishable from zero, and even in a multivariate model each unit of log stars reduces the odds of a high-severity finding by only 22\%, while code size and repository age push the other way.

The contributions are:
\begin{enumerate}
\item a census pipeline and dataset covering the full public MCP registry, its version history, and fetched source for two thirds of the population, with attrition reported rather than imputed;
\item an eight-class threat catalogue operationalized as 31 detection patterns, with per-class precision measured against 414 hand-labeled findings and used to report precision-corrected prevalence;
\item the first population-scale measurement, to our knowledge within this registry, of silent advertised-interface drift and endpoint redirection, including the finding that the registry does not version tool definitions at all;
\item a predictor analysis showing that popularity, maintenance recency, and licensing are weak or null signals of security posture, whereas code size and age are the dominant observable correlates;
\item an anonymized artifact containing the analysis tables, labeling protocol, and scanner configuration.
\end{enumerate}

The remainder of the paper describes background and related work (Section~\ref{sec:related}), the measurement method and its validation (Section~\ref{sec:method}), results per research question (Section~\ref{sec:results}), implications (Section~\ref{sec:discussion}), threats to validity (Section~\ref{sec:threats}), and conclusions (Section~\ref{sec:conclusion}).

\section{Background and Related Work}
\label{sec:related}
This section places the census between two literatures: security analyses of MCP itself, and measurement studies of older package ecosystems whose methods we adapt.

\subsection{MCP and Its Threat Surface}
An MCP server exposes tools, resources, and prompts to a client over stdio or HTTP transports; the client forwards tool descriptions into the model's context and executes the calls the model selects. Hou et al. \cite{hou2025landscape} map this lifecycle and enumerate sixteen threat scenarios across it, from installer spoofing to tool poisoning. Radosevich and Halloran \cite{radosevich2025audit} demonstrate working exploits through MCP-connected models, and Zhao et al. \cite{zhao2025attack} build a taxonomy of ways a server itself can attack its host. Practitioner reports first drew attention to tool poisoning, in which instructions hidden in a tool description manipulate the model rather than the code \cite{invariant2025poisoning}, and to rug-pull dynamics in which a server changes behaviour after gaining trust. Narajala and Habler \cite{narajala2025enterprise} propose mitigation frameworks for enterprise deployment, and the OWASP LLM Top 10 lists supply-chain compromise of tool integrations among its highest-ranked risks \cite{owasp2025llm}.

Empirical work at scale is more recent and thinner. Hasan et al. \cite{hasan2025firstglance} analyse 1{,}899 open-source servers with a hybrid static pipeline and report that 7.2\% contain general vulnerabilities; Guo et al. \cite{guo2025measurement} crawl six third-party marketplaces and find that more than half of listed projects are invalid or low-value; Li and Gao \cite{li2025firstlook} study 67{,}057 entries across six registries with a focus on naming attacks and metadata, identifying 833 vulnerable servers. A separate line systematizes MCP security knowledge \cite{guo2025systematic, gaire2025sok}. Our census differs on three axes. It covers the official registry exhaustively at a snapshot rather than sampling marketplaces; it measures version-to-version drift, which requires the registry's version history and which prior studies do not analyse; and it anchors every prevalence number to hand-labeled precision, so corrected estimates are reported alongside raw ones. Where our populations overlap, the comparison is instructive and we return to it in Section~\ref{sec:discussion}.

\subsection{Measurement of Package Ecosystems}
The questions asked here are the questions the software-engineering community has asked of npm, PyPI, and Maven for a decade. Zimmermann et al. \cite{zimmermann2019smallworld} showed that npm's package graph concentrates risk in a small set of maintainers; Decan et al. \cite{decan2018impact} traced how vulnerabilities propagate through dependency networks; Ohm et al. \cite{ohm2020backstabber} and Ladisa et al. \cite{ladisa2023sok} catalogue real supply-chain attacks, several of which, notably maintainer-account takeover followed by a malicious release, are exactly the dynamic our drift analysis is designed to observe. On the stability side, Raemaekers et al. \cite{raemaekers2014semver} found that Maven releases break interfaces regardless of semantic-versioning promises, and Bogart et al. \cite{bogart2016break} showed that ecosystems differ sharply in their norms around breaking changes. MCP's registry is younger than any of these ecosystems were when first measured, and, as we show, it lacks even the versioning primitives those studies took for granted: tool definitions, the functional interface of a server, are not versioned by the registry at all.

Finally, the measurement instrument itself has a literature. Pattern-based static analysis trades recall for scale and produces false positives whose rate varies by defect class \cite{johnson2013static}; secret-detection studies such as Meli et al. \cite{meli2019git} demonstrated both the value and the noise of regex-based scanning at GitHub scale; and Pearce et al. \cite{pearce2022asleep} used CWE-anchored pattern checks to score generated code, a design similar in spirit to our catalogue. Kalliamvakou et al. \cite{kalliamvakou2014promises} enumerate the pitfalls of mining GitHub, several of which shape our attrition reporting. We follow this literature's central lesson: a scanner's output is a risk indicator, not a verdict, and the honest unit of report is prevalence with measured precision attached.

\section{Method}
\label{sec:method}
The study is a cross-sectional census with a longitudinal component: one registry snapshot supplies the population, its version history supplies the drift measurements, and a hand-labeled sample supplies the error model for the scanner. All parameters described below, including every detection pattern, severity weight, sampling constant, and statistical setting, are fixed in a single configuration file released with the artifact.

\subsection{Pipeline Overview}
Four stages take the registry to the analysis table.

\subsubsection{Registry harvest} We enumerated the full public MCP registry through its paginated API (100 records per page, cursor continuation to exhaustion, 727 pages), capturing every server and every published version at the August 2026 snapshot: 21{,}643 unique servers across 72{,}606 version records, with publication dates spanning September 2025 to August 2026. Per version we retained the description, declared packages with their registries and version constraints (present in 73.1\% of version records), declared remote endpoints with transports (35.9\%), and publication timestamps. Tool definitions appear in none of the 72{,}606 version records because the registry schema does not carry them, a structural fact whose consequences RQ2 measures.

\subsubsection{Source enrichment} For the 17{,}190 servers whose registry entry resolves to a parseable GitHub repository (79.4\%), we fetched repository metadata (stars, language, license, archival status, push and creation timestamps) and source code, with on-disk caching and resumable fetching so that the snapshot is internally consistent. Fetching operated under a fixed budget of 40 files and 200~KB per file per repository, restricted to source and configuration types (Python, JavaScript, TypeScript, JSON, YAML, TOML) plus package and deployment manifests (\texttt{package.json}, \texttt{pyproject.toml}, \texttt{requirements.txt}, \texttt{Dockerfile}, and MCP-specific manifests), prioritizing server entry points and tool handlers, and excluding test, demo, and fixture paths (10{,}004 files skipped by this rule). The result is 13{,}051 repositories, 248{,}817 files, and 1.9~GB of text covering 14{,}353 servers, the \emph{scan population} (66.3\% of the registry). Attrition is reported, not imputed: 2{,}548 servers pointed at repositories that no longer resolve, 263 yielded no files under the budget, 24 had unavailable trees, and single repositories were DMCA-blocked or forbidden. Servers without fetched code are excluded from all denominators rather than assumed clean.

\subsubsection{Scanning} A pattern-based scanner evaluated each fetched file against the threat catalogue of Section~\ref{sec:catalogue} and additionally recovered tool definitions from source, extracting 69{,}027 tool declarations from 4{,}238 repositories. Scanning emitted 4{,}719 findings across 2{,}183 servers.

\subsubsection{Drift comparison} For the 8{,}900 servers with more than one published version, consecutive version pairs (37{,}160 transitions) were differenced on the three fields the registry versions: description text, package identifiers, and remote endpoints. Each detected change event was typed as \emph{additive} (something new is offered alongside the old), \emph{breaking} (a package or endpoint is withdrawn), or \emph{silent semantic} (identifiers persist while meaning or destination changes, as when a description is rewritten or a remote host is replaced). One structural observation frames this stage and recurs throughout the paper: the registry does not version tool definitions, so a change in what a server's tools actually do is invisible in registry data by construction. Drift is therefore measured on the advertised surface, and endpoint redirection serves as the closest observable analogue of a rug pull.

\subsection{Threat Catalogue}
\label{sec:catalogue}
Table~\ref{tab:catalogue} lists the eight classes. They were drawn from the MCP threat literature \cite{hou2025landscape, invariant2025poisoning, narajala2025enterprise} and from classical categories that apply to any network-facing tool server, and each is operationalized as a small set of syntactic patterns (31 in total) over one of three surfaces: recovered tool definitions, handler source code, or package manifests. Severity bands follow the rationale in the table: classes offering a direct path to compromise if reachable are high, classes weakening isolation or confidentiality are medium, and hygiene concerns are low. The severity weights (high 5, medium 3, low 1, informational 0) feed the per-server severity score of \eqref{eq:sev}.

\begin{table}[t]
\caption{Threat catalogue: eight classes, 31 detection patterns.}
\label{tab:catalogue}
\centering
\small
\begin{tabular}{@{}llll@{}}
\toprule
Class & Severity & Surface & Pat. \\
\midrule
Tool poisoning (hidden instructions & high & tool & 4 \\
\quad in tool metadata) & & & \\
Token passthrough / confused deputy & high & code & 5 \\
Command / code injection in handlers & high & code & 5 \\
Unauthenticated network exposure & high & code & 4 \\
Ambient authority (over-broad scope) & medium & code & 4 \\
Insecure secret handling & medium & code & 3 \\
Unvalidated outbound fetch (SSRF) & medium & code & 3 \\
Unpinned or unverified dependency & low & manifest & 3 \\
\bottomrule
\end{tabular}
\end{table}

Each pattern is a guarded regular expression rather than a bare keyword match: guards exclude comments, docstrings, TypeScript type unions, shell keywords, environment-variable names, documented vendor example credentials, and defensive wording (a description warning ``never treat this as an instruction'' must not fire the poisoning detector). The guards raise precision but cannot reach semantics, which is why Section~\ref{sec:validation} measures how far they get per class.

Two classes deserve a word of motivation because they are protocol-specific. \emph{Tool poisoning} scans recovered tool descriptions for embedded instructions aimed at the model rather than the user: overrides of prior instructions, directives to conceal actions from the user, imperatives to read credential files before acting, and directive-bearing emphasis tags of the kind published attacks use \cite{invariant2025poisoning}. \emph{Token passthrough} detects a credential taken from the caller (request parameters or context) being placed into an outbound Authorization header, the confused-deputy pattern \cite{hardy1988confused} in its MCP form: the server acts on authority it did not mint, and any upstream audit trail attributes its actions to the original token holder.

\subsection{Measures and Statistical Analysis}
\label{sec:stats}
All proportions are reported with Wilson score intervals \cite{wilson1927}, which remain calibrated for the rare classes where a normal approximation fails. For $k$ affected servers among $n$ scanned, with $\hat p = k/n$ and $z$ the normal quantile at confidence $1-\alpha$,
\begin{equation}
\label{eq:wilson}
p_{\pm} \;=\; \frac{\hat p + \frac{z^2}{2n} \pm z\sqrt{\frac{\hat p (1-\hat p)}{n} + \frac{z^2}{4n^2}}}{1 + \frac{z^2}{n}} .
\end{equation}

Each server $i$ receives a severity score summing the weights of the classes present,
\begin{equation}
\label{eq:sev}
S_i \;=\; \sum_{c \in \mathcal{C}} w_c \,\mathbf{1}\!\left[k_{i,c} > 0\right],
\qquad w_c \in \{0, 1, 3, 5\},
\end{equation}
where $k_{i,c}$ counts findings of class $c$ in server $i$.

Because the scanner is imperfect, observed prevalence for class $c$ is corrected by the class's measured precision $\hat\rho_c$ from the labeling study of Section~\ref{sec:validation}:
\begin{equation}
\label{eq:corrected}
\tilde\pi_c \;=\; \frac{k_c}{n}\,\hat\rho_c ,
\end{equation}
a deliberately conservative estimator: it removes the false-positive mass but adds nothing back for the (small) measured miss rate, so corrected values are best read as lower bounds on true prevalence. Reported intervals for $\tilde\pi_c$ combine the Wilson interval on the count with the interval on $\hat\rho_c$.

Association between categorical strata uses chi-square tests with Cram\'er's V \cite{cramer1946},
\begin{equation}
\label{eq:cramer}
V \;=\; \sqrt{\frac{\chi^2}{n\,(\min(r, s) - 1)}} ,
\end{equation}
for an $r \times s$ table, alongside odds ratios with 95\% intervals so that effect size is never inferred from $p$-values alone. Finding counts per server are heavy-tailed and zero-inflated, so distributional questions use rank statistics (Spearman, Mann-Whitney, Kruskal-Wallis) with Cliff's delta \cite{cliff1993} where applicable. RQ3's multivariate model is a logistic regression on the probability that server $i$ has any high-severity finding,
\begin{equation}
\label{eq:logit}
\begin{split}
\operatorname{logit} P(H_i = 1) = \beta_0
&+ \beta_1 \log(1{+}\mathrm{stars}_i) \\
&+ \beta_2 \log(1{+}\mathrm{files}_i)
 + \beta_3 \log(1{+}\mathrm{age}_i) \\
&+ \beta_4\,\mathrm{maintained}_i
 + \beta_5\,\mathrm{archived}_i \\
&+ \beta_6\,\mathrm{license}_i ,
\end{split}
\end{equation}
with logs because stars, files, and age each span orders of magnitude, and with code size included because a larger fetched code base gives any pattern more opportunities to fire. Families of related tests are corrected with Benjamini-Hochberg at FDR 0.05 \cite{benjamini1995fdr}; adjusted $p$-values are reported alongside raw ones. Bootstrap intervals use 5{,}000 resamples under a fixed seed (42), and $\alpha = 0.05$ throughout.

\subsection{Scanner Validation}
\label{sec:validation}
A prevalence estimate is only as credible as the detector behind it, so we measured the scanner against manual ground truth before interpreting any population number. From the scanner's output we drew a stratified sample of 294 findings, 42 per class, at most one finding per server, and shuffled the rows so classes were judged in interleaved order, which limits the drift in standards that sets in once a rater notices a pattern. A second stratum of 120 files for which the scanner emitted nothing was inspected to estimate what the scanner misses. In total 414 rows across 389 files were labeled against written per-class decision rules (released with the artifact); the rules resolve genuinely undecidable cases to \emph{false positive}, keeping precision conservative rather than flattering. Labeling was performed by a single rater, a limitation Section~\ref{sec:threats} discusses.

\begin{table}[t]
\caption{Scanner accuracy against 414 manual labels (42 flagged findings per class; 120 unflagged files shared as the negative stratum). Wilson 95\% CIs.}
\label{tab:accuracy}
\centering
\small
\setlength{\tabcolsep}{4.2pt}
\begin{tabular}{@{}lrrrl@{}}
\toprule
Class & TP & FP & Prec. & 95\% CI \\
\midrule
Unpinned dependency & 39 & 3 & 0.93 & [0.81, 0.98] \\
Token passthrough & 33 & 9 & 0.79 & [0.64, 0.88] \\
Unauth.\ exposure & 30 & 12 & 0.71 & [0.56, 0.83] \\
Unvalidated fetch & 30 & 12 & 0.71 & [0.56, 0.83] \\
Insecure secrets & 16 & 26 & 0.38 & [0.25, 0.53] \\
Ambient authority & 10 & 32 & 0.24 & [0.13, 0.39] \\
Command injection & 7 & 35 & 0.17 & [0.08, 0.31] \\
\midrule
Pooled & 165 & 129 & 0.56 & [0.50, 0.62] \\
\bottomrule
\end{tabular}
\end{table}

\begin{figure}[t]
\centering
\includegraphics[width=0.62\textwidth]{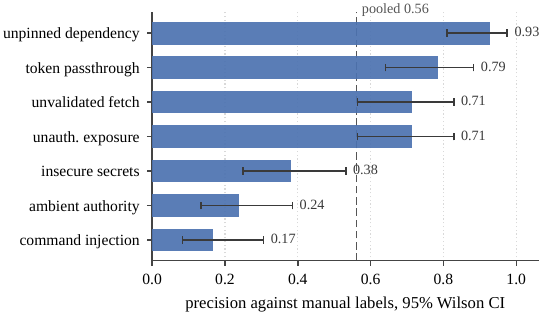}
\caption{Per-class precision against manual labels, Wilson 95\% CIs; the dashed line marks pooled precision (0.56). Narrow syntactic signatures score high; use-dependent classes score low.}
\label{fig:precision}
\end{figure}

Table~\ref{tab:accuracy} and Fig.~\ref{fig:precision} give the outcome. Pooled precision is 0.56 (95\% CI [0.50, 0.62]), and the spread across classes is wide and systematic: classes defined by a narrow syntactic signature (an unconstrained version specifier, a caller token in an Authorization header) reach precision of 0.79 to 0.93, while classes whose truth depends on how a value is later used (whether a string reaches a shell, whether a wildcard scope is enforced) fall to 0.17 to 0.24. This ordering is the expected behaviour of pattern-based analysis \cite{johnson2013static} and it defines the boundary of what the method can claim; it is also exactly why \eqref{eq:corrected} feeds a per-class rather than pooled correction into the results. Recall within the labeled sample is high (0.94 to 1.00 per class), and of the 120 unflagged files, 2 contained an issue a rater judged real (both token-passthrough cases), a miss rate of 0.017 (95\% CI [0.005, 0.059]). Population recall is not identifiable because the true number of issues in unscanned code is unknown; the miss rate instead bounds how much the reported prevalences understate the ecosystem.

\subsection{Ethics}
\label{sec:ethics}
All collected data are public: registry metadata, public repositories, and repository statistics. Findings are risk indicators from static patterns, not verified exploits, and we neither executed servers nor probed any deployed endpoint. Results are reported in aggregate throughout; no server, maintainer, or repository is named, and the released artifact replaces server identities in finding-level data with salted hashes and truncates evidence strings so that the artifact cannot serve as a target list. The insecure-secrets class surfaced a small number of credential values that appear live rather than placeholders; prior to submission we reported these through the hosting platform's private disclosure channels and to the issuing providers' credential-revocation programs, and we withhold the affected rows from the public artifact until remediation. The census imposed negligible load on public APIs (rate-limited, cached, resumable fetching over roughly eight hours).
\section{Results}
\label{sec:results}
This section reports the three research questions in turn; every estimate carries a 95\% interval, and prevalence denominators are always the scan population ($n = 14{,}353$) unless stated otherwise. Table~\ref{tab:corpus} summarizes the corpus each analysis draws on. Two properties of the population are worth fixing in mind before any prevalence figure: it is young and it is obscure. The earliest version record in the registry dates to September 2025, eleven months before the snapshot; the median server has one star (7{,}156 servers have zero); the median repository was pushed 16 days before the snapshot; and 41\% of servers have published more than one version. The census therefore describes an ecosystem in its formative period, which is precisely when structural fixes are cheapest.

\begin{table}[t]
\caption{Corpus construction. Percentages are of the registry population; attrition is reported, not imputed.}
\label{tab:corpus}
\centering
\small
\begin{tabular}{@{}lr@{}}
\toprule
Stage & Count \\
\midrule
Servers in registry (Aug.\ 2026 snapshot) & 21{,}643 \\
Version records & 72{,}606 \\
Servers declaring a repository (79.4\%) & 17{,}190 \\
Repositories fetched with files & 13{,}051 \\
Source files scanned (1.9 GB) & 248{,}817 \\
Test/demo/fixture files excluded & 10{,}004 \\
\textbf{Scan population} (66.3\%) & \textbf{14{,}353} \\
Multi-version servers (RQ2 set) & 8{,}900 \\
Tool definitions recovered & 69{,}027 \\
Findings emitted & 4{,}719 \\
Hand-labeled rows (validation) & 414 \\
\bottomrule
\end{tabular}
\end{table}

\subsection{RQ1: Threat-Class Prevalence}
\begin{figure}[t]
\centering
\includegraphics[width=0.62\textwidth]{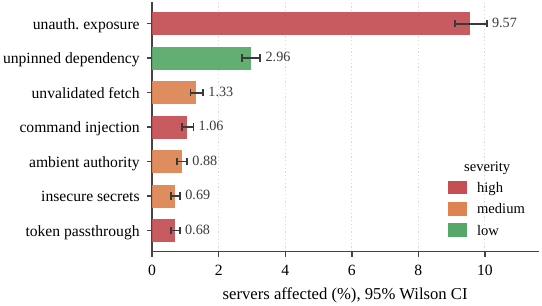}
\caption{Observed prevalence per threat class with Wilson 95\% CIs ($n = 14{,}353$ scanned servers); bar color encodes the severity band.}
\label{fig:prevalence}
\end{figure}

\begin{figure}[t]
\centering
\includegraphics[width=0.62\textwidth]{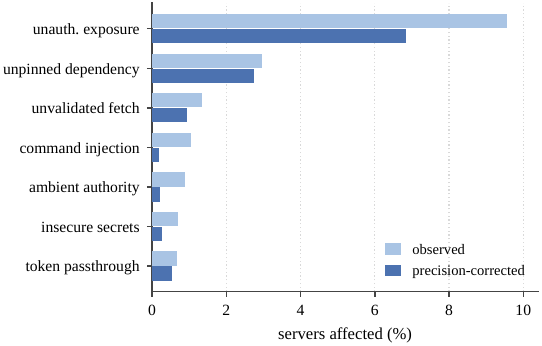}
\caption{Observed versus precision-corrected prevalence per class. Correction reorders the classes: apparent command-injection and ambient-authority prevalence is mostly scanner noise, while exposure and unpinned dependencies survive nearly intact.}
\label{fig:corrected}
\end{figure}

The headline is concentration: findings are common, but one class carries most of the weight. Across the scan population, 2{,}183 servers (15.21\%, CI \ci{14.63}{15.81}) carry at least one finding and 1{,}599 (11.14\%, CI \ci{10.64}{11.67}) at least one high-severity finding. Table~\ref{tab:prevalence} and Fig.~\ref{fig:prevalence} break this down by class, and Fig.~\ref{fig:corrected} shows the same classes after the per-class precision correction of \eqref{eq:corrected}.

\begin{table}[t]
\caption{Observed and precision-corrected prevalence per class (\% of $n = 14{,}353$ scanned servers, 95\% CIs).}
\label{tab:prevalence}
\centering
\small
\setlength{\tabcolsep}{3.4pt}
\begin{tabular}{@{}lrrlrl@{}}
\toprule
Class & $k$ & Obs. & 95\% CI & Corr. & 95\% CI \\
\midrule
Unauth.\ exposure & 1{,}373 & 9.57 & [9.10, 10.06] & 6.83 & [5.40, 7.92] \\
Unpinned dep. & 425 & 2.96 & [2.70, 3.25] & 2.75 & [2.40, 2.89] \\
Unvalidated fetch & 191 & 1.33 & [1.16, 1.53] & 0.95 & [0.75, 1.10] \\
Command injection & 152 & 1.06 & [0.90, 1.24] & 0.18 & [0.09, 0.32] \\
Ambient authority & 127 & 0.88 & [0.74, 1.05] & 0.21 & [0.12, 0.34] \\
Insecure secrets & 99 & 0.69 & [0.57, 0.84] & 0.26 & [0.17, 0.37] \\
Token passthrough & 98 & 0.68 & [0.56, 0.83] & 0.54 & [0.44, 0.60] \\
Tool poisoning & 0 & 0.00 & [0.00, 0.03] & -- & -- \\
\bottomrule
\end{tabular}
\end{table}

Three observations follow. First, unauthenticated network exposure dominates: binding to all interfaces, disabling authentication, or setting a wildcard CORS origin affects nearly one server in ten, and at 0.71 precision most of that mass is real (corrected 6.83\%). Because the classes rarely co-occur (84.8\% of servers have none, 13.3\% exactly one, 1.9\% two, 0.06\% three), composing the corrected high-severity classes gives roughly 7.6\% as a working estimate of true high-severity prevalence, against 11.1\% observed. Second, the correction reorders the ranking. Command injection appears at 1.06\% but corrects to 0.18\%, and ambient authority from 0.88\% to 0.21\%, because those detectors have precision of 0.17 and 0.24; any reader comparing raw scanner output across studies should expect this effect. Third, tool poisoning, the most discussed MCP-specific attack, was not detected at all: zero of 14{,}353 servers, with a Wilson upper bound of 0.027\%. Absence of detection is not proof of absence, since pattern matching cannot see obfuscated variants, but it does mean overt poisoning of the kind published in demonstrations \cite{invariant2025poisoning} is not present at measurable rates in the registry population, while the mundane classes are.

Prevalence also varies by how a server is delivered. High-severity prevalence differs across transports ($\chi^2(2) = 375.04$, $p < .001$ raw and BH-adjusted, $V = 0.16$): servers declaring the deprecated SSE transport reach 44.2\% (136 of 308) against 12.3\% for streamable HTTP and 9.3\% for servers with no remote, consistent with SSE marking older, less maintained code. Differences across package registries are detectable but small ($\chi^2(4) = 141.35$, $p < .001$, $V = 0.10$), with pypi- and mcpb-delivered servers around 16\% to 17\% and npm at 8.0\%.

\subsubsection{Delegation Risk}
Ambient authority and token passthrough warrant joint treatment: both concern authority the server did not create, one by granting itself over-broad scope, the other by forwarding a caller's credential upstream. Either class affects 225 servers (1.57\%, CI \ci{1.38}{1.78}); notably the two never co-occur (0 of 14{,}353, upper bound 0.03\%), so they mark disjoint failure modes rather than a shared style of careless code. Corrected, the joint figure is roughly 0.75\% (0.21\% ambient authority, 0.54\% token passthrough), and token passthrough is the more trustworthy signal at 0.79 precision. Delegation risk concentrates where it matters most: remote-hosted servers show 2.09\% prevalence against 1.25\% for locally installed ones (OR = 1.69, CI \ci{1.30}{2.20}, $\chi^2(1) = 15.01$, $p < .001$ adjusted), and a remote host holds any forwarded credential on infrastructure the user does not control, so the same code pattern carries more consequence there.

\subsection{RQ2: Silent Drift of the Advertised Interface}
\begin{figure}[t]
\centering
\includegraphics[width=0.62\textwidth]{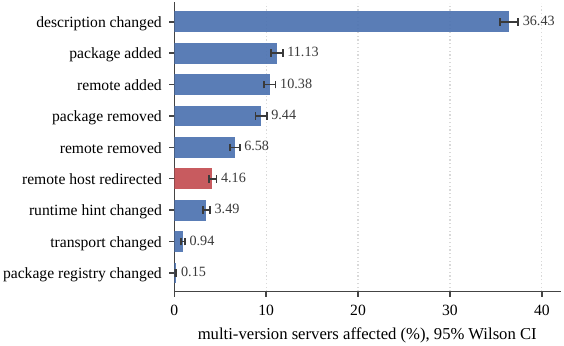}
\caption{Share of multi-version servers affected by each change kind, Wilson 95\% CIs ($n = 8{,}900$). Endpoint host redirection (highlighted) is the closest observable analogue of a rug pull.}
\label{fig:driftkinds}
\end{figure}

\begin{figure}[t]
\centering
\includegraphics[width=0.62\textwidth]{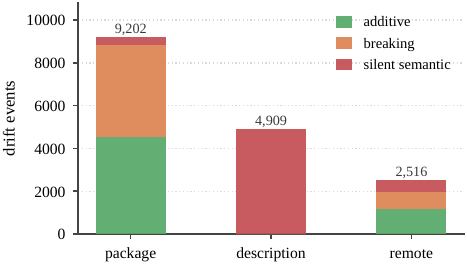}
\caption{Drift events by surface and drift type. Package fields account for most events; description changes are silent-semantic by construction, and silent changes appear on every surface.}
\label{fig:surface}
\end{figure}

What a registry name resolves to turns out to be unstable, and the instability is mostly invisible to clients. Among the 8{,}900 multi-version servers, 37{,}160 version transitions yielded 16{,}627 change events. Just over half of these servers, 4{,}547 (51.09\%, CI \ci{50.05}{52.13}), changed description, packages, or endpoints across versions; 40.58\% (CI \ci{39.57}{41.61}) had at least one \emph{silent semantic} change, in which every identifier stayed fixed while meaning or destination moved; 20.96\% had additive and 15.67\% breaking changes. Fig.~\ref{fig:driftkinds} and Table~\ref{tab:driftkinds} decompose the changes by kind, and Fig.~\ref{fig:surface} by surface: package fields generate most events (9{,}202), then descriptions (4{,}909), then remotes (2{,}516).

\begin{table}[t]
\caption{Change kinds among multi-version servers ($n = 8{,}900$; a server can appear in several rows).}
\label{tab:driftkinds}
\centering
\small
\begin{tabular}{@{}lrrl@{}}
\toprule
Change kind & Servers & \% & 95\% CI \\
\midrule
Description changed & 3{,}242 & 36.43 & [35.43, 37.43] \\
Package added & 991 & 11.13 & [10.50, 11.81] \\
Remote added & 924 & 10.38 & [9.77, 11.03] \\
Package removed & 840 & 9.44 & [8.85, 10.06] \\
Remote removed & 586 & 6.58 & [6.09, 7.12] \\
Remote host redirected & 370 & 4.16 & [3.76, 4.59] \\
Runtime hint changed & 311 & 3.49 & [3.13, 3.90] \\
Transport changed & 84 & 0.94 & [0.76, 1.17] \\
Package registry changed & 13 & 0.15 & [0.09, 0.25] \\
\bottomrule
\end{tabular}
\end{table}

The most security-relevant kind is endpoint redirection: 370 servers (4.16\%, CI \ci{3.76}{4.59}; 441 events) kept their registry identity while moving their declared remote endpoint to a different host. For an already-installed client this is the observable shape of a rug pull: the same name resolves, traffic reaches a different origin, and the protocol provides no notification of the change. We do not claim these 370 redirections are malicious; most are surely routine migrations. The point is structural: the registry cannot distinguish a migration from a takeover, and neither can a client. A robustness check matters here because a few servers publish many versions per day (one published 1{,}159): capping the analysis at 20 transitions per server moved the redirection estimate by less than 0.05 percentage points, so the finding is not an artifact of bulk republishers.

Drift also moves fast. Among servers that changed at all, the median time to first change is 0 days (mean 10.3, SD 27.0, Q3 6.5), and the median gap between changed versions is one day. Whatever review a user performed at install time describes, at the median, a server that had already changed the same day.

Finally, instability correlates with security posture. Restricting to the 6{,}699 multi-version servers inside the scan population, servers with silent drift have 2.96 times the odds of a high-severity finding (CI \ci{2.56}{3.42}, $\chi^2(1) = 231.98$, $p < .001$ raw and adjusted, $V = 0.19$) and 2.34 times the odds of any finding (CI \ci{2.06}{2.65}, $\chi^2(1) = 177.51$, $p < .001$, $V = 0.16$). The design is correlational and the direction is not identified; the parsimonious reading is that both quantities proxy the same latent variable, low process maturity, which makes advertised-interface churn a cheap, registry-observable warning sign for weaknesses that otherwise require source access to detect.

\subsection{RQ3: What Predicts Security Posture}
If users cannot audit code, they fall back on visible signals, so it matters which signals carry information. Popularity carries almost none. Spearman correlations between stars and finding counts ($\rho = 0.008$, $p = .335$, $n = 14{,}353$) and between stars and severity score ($\rho = 0.010$, adjusted $p = .308$) are indistinguishable from zero. Recency of maintenance is similarly thin: days since last push correlates weakly and negatively with findings ($\rho = -0.033$, adjusted $p < .001$), an effect too small to act on, and the bivariate comparison of maintained (pushed within 180 days) versus stale servers finds nothing (OR = 1.04, CI \ci{0.77}{1.41}, $p = .857$). Repository age is the strongest simple correlate ($\rho = 0.087$, adjusted $p < .001$): older servers carry more findings.

\begin{figure}[t]
\centering
\includegraphics[width=0.62\textwidth]{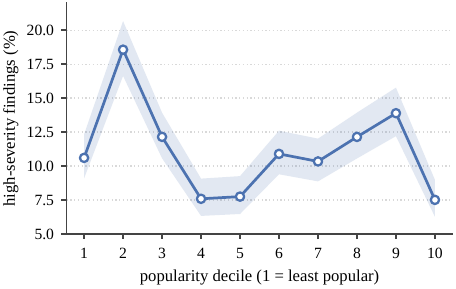}
\caption{High-severity prevalence by popularity decile with 95\% CIs. The relation is non-monotone; decile 2 peaks at 18.6\% and the top decile is lowest at 7.5\%.}
\label{fig:decile}
\end{figure}

Fig.~\ref{fig:decile} shows why the rank correlation is null: the relation is non-monotone. High-severity prevalence peaks in decile 2 (18.57\%, CI \ci{16.63}{20.67}), dips in the mid-deciles, rises again through deciles 8 and 9, and only drops at the very top (decile 10: 7.51\%, CI \ci{6.27}{8.98}). Comparing extremes, the most popular decile does beat the least popular (OR = 0.68, CI \ci{0.53}{0.89}, $\chi^2(1) = 7.90$, $p = .005$, $V = 0.05$), but the effect is small and everything between the extremes is unordered.

\begin{table}[t]
\caption{Logistic regression predicting any high-severity finding ($n = 14{,}353$, pseudo-$R^2 = 0.050$). Odds ratios per unit of the (log-transformed) predictor.}
\label{tab:logit}
\centering
\small
\begin{tabular}{@{}lrlr@{}}
\toprule
Term & OR & 95\% CI & $p$ \\
\midrule
log(1+stars) & 0.78 & [0.75, 0.82] & $<.001$ \\
log(1+files) & 1.94 & [1.81, 2.09] & $<.001$ \\
log(1+age) & 1.38 & [1.29, 1.48] & $<.001$ \\
Maintained (180 d) & 1.22 & [0.89, 1.68] & .218 \\
Archived & 1.52 & [1.05, 2.21] & .027 \\
Has license & 1.01 & [0.86, 1.19] & .859 \\
\bottomrule
\end{tabular}
\end{table}

\begin{figure}[t]
\centering
\includegraphics[width=0.62\textwidth]{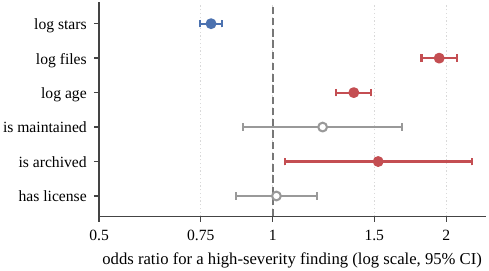}
\caption{Odds ratios with 95\% CIs from the logistic model of \eqref{eq:logit}. Filled markers: CI excludes 1 (red raises odds, blue lowers); open gray markers: no reliable effect.}
\label{fig:logit}
\end{figure}

The multivariate model (Table~\ref{tab:logit}, Fig.~\ref{fig:logit}) separates the confounded pieces. Holding size and age fixed, popularity is protective but modestly so: each unit of log stars multiplies the odds of a high-severity finding by 0.78 (CI \ci{0.75}{0.82}), meaning a 1{,}000-star server has roughly a fifth the odds of an otherwise identical zero-star server, but ``otherwise identical'' is doing real work because popular servers are also larger, and each unit of log files nearly doubles the odds (OR = 1.94, CI \ci{1.81}{2.09}). Age adds risk (OR = 1.38), archived repositories add risk (OR = 1.52, $p = .027$), and neither recent maintenance nor a declared license carries signal once the structural variables are in the model. The model's pseudo-$R^2$ of 0.050 is itself a finding: the metadata a user can see explains little of where weaknesses sit, which undercuts any install-time heuristic built on stars, freshness, or licensing.

\section{Discussion}
\label{sec:discussion}
The census supports one overall reading: the acute problem in the MCP ecosystem today is not exotic attacks but a registry that cannot describe change, layered over deployment defaults that fail open. This section draws out what follows for registry operators, client developers, security tooling, and researchers.

\subsection{The Registry Should Version What Matters}
The single most consequential structural fact is that tool definitions, the functional contract a server presents to a model, are not versioned by the registry, so a behavioural change is undetectable from registry data by construction. Even on the fields the registry does version, 40.6\% of multi-version servers changed meaning or destination under stable identifiers, and 4.2\% redirected their endpoint host with no mechanism to inform installed clients. Package ecosystems learned this lesson expensively: attacks catalogued from Ohm et al. \cite{ohm2020backstabber} to Ladisa et al. \cite{ladisa2023sok} repeatedly exploit the gap between a stable name and mutable content, and registries responded with immutable versions, provenance attestation, and yanked-release signalling. Concrete steps for MCP follow the same path: version tool definitions in the registry; treat a remote-endpoint host change as a breaking release requiring re-consent; and expose a machine-readable diff between versions so clients can alert rather than silently follow. Until then, clients can act unilaterally with trust-on-first-use pinning of endpoint hosts and package digests, surfacing any deviation to the user.

\subsection{Defaults, Not Attackers, Drive Prevalence}
Corrected prevalence says the ecosystem's weight problem is unauthenticated exposure (6.8\%) and unpinned dependencies (2.8\%), not injected shells or poisoned metadata. The exposure evidence is dominated by three recurring lines: binding to all interfaces, wildcard CORS, and explicitly disabled authentication, which are the path of least resistance in current server templates and frameworks. The cheapest ecosystem-wide risk reduction is therefore in framework defaults: bind loopback unless configured otherwise, refuse wildcard origins outside a development flag, and require an explicit opt-out for authentication. The same logic that made memory-safe defaults more productive than exploit-by-exploit patching applies here at the configuration layer.

\subsection{Scanner Triage and Honest Reporting}
The validation numbers carry a methodological message beyond this paper. Raw pattern-scanner output overstated command injection by a factor of six and ambient authority by a factor of four, while barely overstating dependency and passthrough findings. Studies that report raw static-analysis prevalence without measuring per-class precision are therefore not comparable to each other, and differences between published MCP prevalence figures, such as the 7.2\% vulnerability rate of Hasan et al. \cite{hasan2025firstglance} on 1{,}899 servers or the 833 vulnerable servers of Li and Gao \cite{li2025firstlook} across six registries, may reflect detector calibration as much as population differences. For practitioners the per-class ordering is directly actionable: high-precision classes (unpinned dependencies, token passthrough) can gate CI with tolerable noise, while low-precision classes (command injection, ambient authority) should trigger data-flow confirmation before a human sees them.

\subsection{Popularity Is Not a Safety Signal}
Users and marketplaces routinely rank servers by stars. The census gives that heuristic little support: rank correlations are null, the decile curve is non-monotone with its peak just above the bottom, and the multivariate protective effect of popularity is modest against the opposing pull of code size. A plausible mechanism for the decile-2 peak is servers with just enough visibility to attract deployment but not enough scrutiny to attract review; whatever the mechanism, install-time trust decisions need registry-side evidence (versioned interfaces, provenance, scan attestations) rather than social proof.

\section{Threats to Validity}
\label{sec:threats}
\subsubsection{Construct validity} Findings are pattern-based risk indicators, not confirmed vulnerabilities; the scanner cannot establish reachability or exploitability. We mitigate by measuring per-class precision on 414 labels and reporting corrected prevalence, but \eqref{eq:corrected} corrects only the false-positive direction, so corrected values are lower bounds up to the measured 1.7\% miss rate. The zero detection of tool poisoning bounds only overt, pattern-visible poisoning; obfuscated instructions in metadata would evade the detector. Drift is measured on the fields the registry versions, so behavioural change below that surface is invisible, which understates instability rather than overstating it. The 40-file, 200~KB per-repository budget can miss findings in very large repositories, again in the conservative direction.

\subsubsection{Internal validity} Ground truth comes from a single rater; the protocol counters this with written per-class rules, interleaved presentation, and a rule resolving undecidable cases against the scanner, but inter-rater agreement could not be computed and the precision estimates inherit single-rater subjectivity. All associations (drift with findings, predictors with posture) are cross-sectional and correlational; no causal direction is claimed.

\subsubsection{External validity} The population is the official public registry at one snapshot. Third-party marketplaces list servers we do not cover, and results may not transfer to them \cite{guo2025measurement, li2025firstlook}. Within the registry, 33.7\% of servers lack fetched code, mostly for want of a resolvable repository, and these are excluded from denominators; if closed-source servers differ systematically, prevalence for the full registry differs from the scan population. GitHub-hosted metadata carries the known biases of mining GitHub \cite{kalliamvakou2014promises}.

\subsubsection{Conclusion validity} Rare-class intervals use Wilson scores, related test families are BH-corrected at FDR 0.05, effect sizes accompany every test, and the bootstrap and sampling seeds are fixed, so the numbers are reproducible from the artifact.

\section{Conclusion}
\label{sec:conclusion}
A census of 21{,}643 MCP servers finds an ecosystem whose weaknesses are mundane and whose instability is structural. One scanned server in nine exposes an unauthenticated surface, and the corrected high-severity rate near 7.6\% is driven by fail-open defaults rather than sophisticated attacks. Meanwhile half of multi-version servers changed what they advertise, two in five silently, 370 moved their endpoint host under a stable name, and silent drift triples the odds of a high-severity finding, all against a registry that does not version the tool definitions clients depend on. Popularity, freshness, and licensing tell a user almost nothing. The path forward is registry-level: version the functional interface, make endpoint changes consensual, and give clients a diff to react to. The dataset and pipeline released with this paper provide the baseline against which those changes can be measured.

\bibliographystyle{unsrt}
\bibliography{references}

\end{document}